# Classical and modern power spectrum estimation for tune measurement in CSNS RCS

Yang Xiao-yu(杨晓宇)    Xu Tao-guang(徐韬光)    Fu Shi-nian(傅世年)
Zeng-lei(曾磊)    Bian Xiao-juan(边晓娟)

Institute of High Energy Physics, CAS, Beijing 100049, China

**Abstract:** Precise measurement of betatron tune is required for good operating condition of CSNS RCS. The fractional part of betatron tune is important and it can be measured by analyzing the signals of beam position from the appointed BPM. Usually these signals are contaminated during the acquisition process, therefore several power spectrum methods are used to improve the frequency resolution. In this article classical and modern power spectrum methods are used. In order to compare their performance, the results of simulation data and IQT data from J-PARC RCS are discussed. It is shown that modern power spectrum estimation has better performance than the classical ones, though the calculation is more complex.

**Introduction**

China Spallation Neutron Source (CSNS) Rapid Cycling Synchrotron (RCS) is designed to accelerate the proton beam to 1.6GeV at 25Hz repetition rate, during this process, the revolution frequency of beam is changed from 0.51MHz to 1.21MHz. The design tune $\nu_x$ is 4.86 and $\nu_y$ is 4.78, here $\nu_{x,y}$ is the ratio of the frequency of betatron oscillation to the revolution frequency of the bunch in transverse direction[1].

For betatron tune measurement in CSNS RCS, beam is planned to be excited by the stripline kicker fed with white noise, and beam position signals from the specified beam position monitor (BPM) will be digitized turn by turn or continuously. In frequency domain the spectrum of betatron oscillation is seen as sidebands on both sides of the corresponding revolution harmonic, as the carrier wave is modulated by betatron oscillation.

The role of power spectrum estimation in random signal analysis is similar to that of Fourier spectrum in determinate signal. It mainly includes classical and modern spectrum estimation methods. Classical methods are nonparametric methods, in which the estimation of power spectrum is made directly from the signal itself. The simplest method is the Periodogram method[2], and an improved method of Periodogram is the Welch method[3].

Modern spectrum method here discussed in our article is a parametric method. The power spectrum is estimated by first estimating the parameters of the linear system that hypothetically generates the signal, as the power spectrum is assumed to be the output of a linear system driven by white noise. The Burg method is such a method and will be discussed later. Modern spectrum methods tend to produce better results than classical nonparametric methods when we adopt the appropriate parametric model[4].

**Periodogram method**

The Periodogram method is based on a Fourier series model of the data, having direct access to Fourier transform $X_N(e^{j\omega})$ of $N-\text{point}$ observation data $x_N(n)$. The power spectrum is estimated by making the square the amplitude of $X_N(e^{j\omega})$ divided by $N$.

$$\hat{P}_{per}(e^{j\omega}) = \frac{1}{N}\left|\sum_{0}^{N-1} x(n)e^{-j\omega n}\right|^2 \tag{1}$$

Periodogram is calculated efficiently and a reasonable result can be produced for a large set of data[5]. In spite of these advantages, there are a few inherent performance deficiencies of this method. The most prominent one is the spectrum leakage, resulting in the energy in main lobe of a spectrum leaks into the side lobes. The reason for this is the truncation effect, due to the assumption that the data outside observation are defaulted as zeros. It equals the result multiplying the data by a rectangular window in time domain, which breaks the correlation of the data inside and outside observation. There are two parameters to indicate the performance of window function, main lobe width and side lobe level. Rectangular window has the narrowest main lobe, corresponding to the best resolution, but the most serious leak occurs for the bad side lobe level of rectangular window. Narrow main lobe and low side lobe level are expected, but they are a couple of contradictory parameters. Hamming window and Hann window are usually adopted for the balance between these two parameters. Basic parameters of four types of window function with equal length 1024 are compared in Table 1.

Table 1: Basic parameters of four types of window function

| Window function | Performance of spectrum of window function | | |
|---|---|---|---|
| | Main lobe width (* π rad/sample) | Relative side lobe attenuation (dB) | Leakage factor / % |
| Rectangular window | 0.001709 | -13.3 | 9.15 |
| Triangular window | 0.002441 | -26.5 | 0.28 |
| Hann window | 0.002686 | -31.5 | 0.05 |
| Hamming window | 0.002441 | -42.7 | 0.03 |

**Welch method**

The Welch method is a modification of Periodogram. $N-po\text{int}$ observation data $x_N(n)$ is divided into $M$ overlapping or nonoverlapping segments to reduce large variance of Periodogram. Assuming there are $L-po\text{int}$ data in each segment, the appointed window is applied to each segment to reduce side lobe effect. The modified Periodogram of the $i\ th$ segment is given by

$$\hat{P}_{per}^{i}(e^{j\omega}) = \frac{1}{U}\left|\sum_{n=0}^{L-1} x_i(n)\omega(n)e^{-j\omega n}\right|^2 \quad i=1,\ 2,\ 3,\ \cdots,\ M \tag{2}$$

where $U$ is called normalization factor, and it is given by

$$U = \frac{1}{L}\sum_{n=0}^{L-1}\omega^2(n) \tag{3}$$

Based on the modified Periodogram of each segment, the power spectrum of $x_N(n)$ can be estimated by averaging $M$ modified Periodogram

$$\hat{P}_{Welch} = \frac{1}{M}\sum_{i=1}^{M}\hat{P}_{per}^{i}(e^{j\omega}) \tag{4}$$

It can be extended as

$$\hat{P}_{Welch} = \frac{1}{MU}\sum_{i=1}^{M}\left|\sum_{n=0}^{L-1}x_i(n)\omega(n)e^{-j\omega n}\right|^2 \tag{5}$$

When the length of $x_N(n)$ is fixed, with the increase of the number of the segments, the length of data which each segment contains will decrease[6]. It is helpful to decrease the variance of estimation by using a large number of segments, but the frequency resolution of power spectrum will deteriorate due to a small quantity of data. Therefore the choice of the number of segments should be considered according to the requirement.

**Burg method**

The Periodogram method and Welch method are nonparametric methods, while the Burg method is a parametric method[7]. For parametric method, the estimation is based on the parametric model of random process. There are three types of model, Autoregressive Recursive (AR) model, Moving Average (MA) model and Autoregressive Recursive Moving Average (ARMA) model. Compared with ARMA and MA model, there is a significant advantage for the AR model, which has no need to solve nonlinear equation for parameter estimation, therefore the AR model is more popular in engineering application[8].

The Burg method for AR power spectrum is based on minimizing the forward and backward predication errors while satisfying the Levinson-Durbin recursion. It avoids calculating the autocorrelation function, while estimates the AR parameters directly instead.

For $N-point$ observation data $x_N(n)$, the $p\,th$ forward and backward predication errors can be defined as

$$e_p^f(n) = x(n) + \sum_{k=1}^{p}a_{pk}x(n-k) \tag{6}$$

$$e_p^b(n) = x(n-p) + \sum_{k=1}^{p}a_{pk}^{*}x(n-p+k) \tag{7}$$

The average power of the $p\,th$ forward and backward predication errors can be defined as

$$\rho_{p,f} = \frac{1}{N-p}\sum_{n=p}^{N-1}\left|e_p^f(n)\right|^2 \tag{8}$$

$$\rho_{p,b} = \frac{1}{N-p}\sum_{n=p}^{N-1}\left|e_p^b(n)\right|^2 \tag{9}$$

The average power of the $p\,th$ predication errors can be given by

$$\rho_p = \frac{1}{2}(\rho_{p,f} + \rho_{p,b}) \qquad (10)$$

And the $p\,th$ reflection coefficient $k_p$ can be calculated by minimizing the average power of predication errors, making $\dfrac{\partial \rho_p}{\partial k_p} = 0$

$$k_p = \frac{-2\sum_{n=p}^{N-1} e_{p-1}^{f}(n) e_{p-1}^{b*}(n-1)}{\sum_{n=p}^{N-1}\left(\left|e_{p-1}^{f}(n)\right|^2 + \left|e_{p-1}^{b}(n-1)\right|^2\right)} \qquad (11)$$

And the corresponding model parameters can be calculated by

$$a_{p,i} = a_{p-1,i} + k_p a_{p-1,p-i} \qquad i = 1, 2, \cdots, p-1 \qquad (12)$$

$$a_{p,p} = k_p \qquad (13)$$

The power spectrum of $x_N(n)$ can be estimated from these model parameters, and the Burg method can always produce a stable model. But the accuracy is lower for high-order model, therefore the adoption of order is important for the Burg method.

**Experiments with simulation data**

Frequency resolution is an important indicator for evaluating the performance of power spectrum. Here two kinds of situations are discussed, one is the distinction of two signals when their frequencies are very close, and the other is the detection of the split of spectrum peak when Signal to Noise Ratio (SNR) is low[9].

In the first situation, the frequencies of two signals are 330Hz and 335Hz, and their amplitudes are equal in time domain. The sampling rate is 3000Hz and the number of samples is 1024. Three power spectrum methods discussed above are used to distinguish these two signals. For the Welch method, hamming window is used, the length we choose is 512, and the overlap is 400. For the Burg method, 600 order parameter model is adopted. In Fig.1 it is shown that these two signals can not be distinguished by using the Periodogram method and the Welch method, because there is only one peak, and the frequency is 334Hz for both of the two methods. For the Burg method, two spectrum peaks with frequencies 328.1Hz and 336.9Hz are seen, and their amplitudes are similar. It shows that the Burg method has higher frequency resolution than classical power spectrum methods.

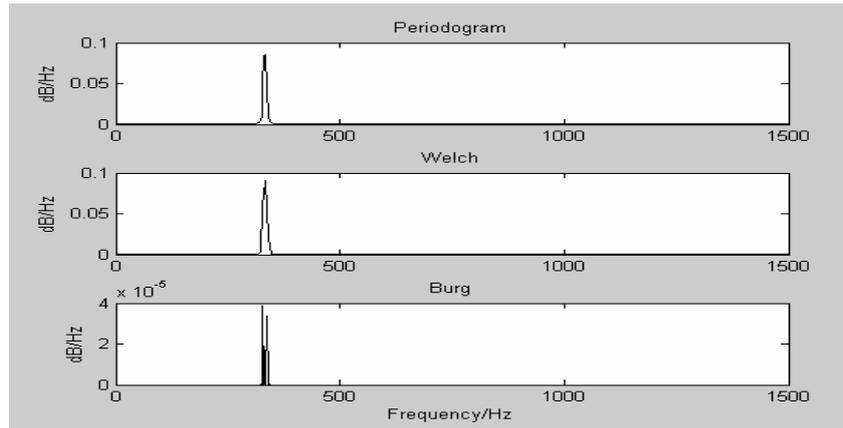

Fig.1 The distinction of two signals with close frequencies

As betatron oscillation is seen as sidebands on both sides of the corresponding revolution harmonic in frequency domain, one amplitude modulated wave is created to simulate this situation. Here the frequency of carrier wave is 300Hz and the frequency of modulating wave is 30Hz, corresponding to the frequencies of two sidebands which are 270Hz and 330Hz. In the first subplot of Fig.2, the power spectrum of signal without noise calculated by using the Periodogram method is shown, the frequencies of two sidebands are 269.5Hz and 331.1Hz. In the other three subplots, white noise is added to the original wave, here SNR -6.217dB is considered. Three power spectrum methods are compared and the frequencies of sidebands are calculated. In the second subplot, for the Periodogram method, the frequency of lower sideband is 269.5Hz but there is some split of spectrum peak occurring in the upper sideband, corresponding to two frequency components, 328.1Hz and 334Hz, therefore it is difficult to find the upper sideband and difficult to calculate the tune value. In the third subplot, for the Welch method, two obvious sidebands are seen. The frequency of lower sideband is 269.5Hz, while the upper sideband is 328.1Hz, but the Full Width at Half Maxium (FWHM) of sidebands become wider, it means the frequency resolution has deteriorated[10]. For the last subplot, the modern power spectrum, the Burg method, the lower sideband and upper sideband are obvious and thinner than the Welch method, their frequencies are 269.5Hz and 331.1Hz, the same as the original signal. By comparison, it is known that the Burg method can identify the characteristic of signal while keeping higher frequency resolution, also it hardly ever generates false peak and this is useful for tune measurement when the SNR is low.

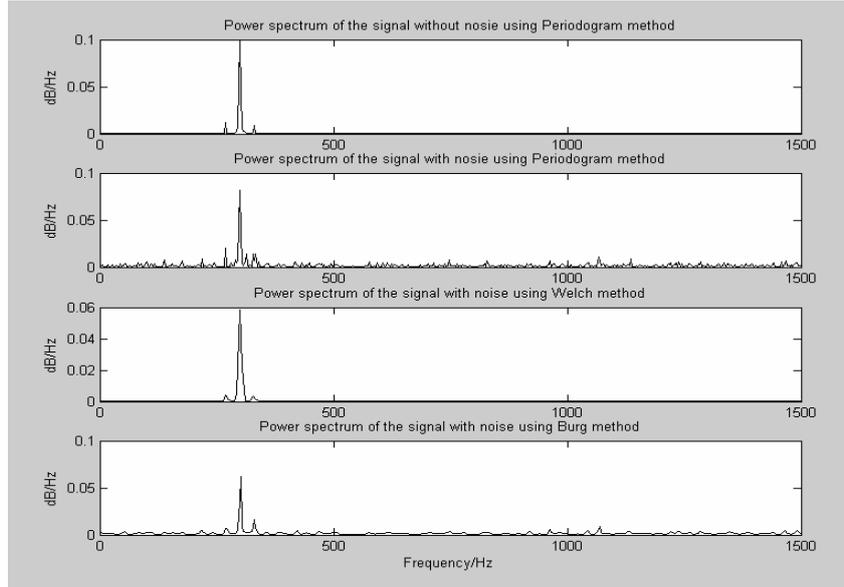

Fig.2 The detection of split of spectrum peak when the SNR is low

**Experiments with experiment data from J-PARC RCS**

As CSNS RCS has not been built yet and the main parameters of Japan Proton Accelerator Research Complex (J-PARC) RCS are similar to CSNS RCS, we analyze the IQT data file from Tektronix spectrum analyzer in J-PARC RCS to compare the performance of classical and modern power spectrum methods for tune measurement[11].

IQT data is one type of data format used for saving BPM data in J-PARC RCS. Usually one file contains 156 frames and each frame contains 1024 pairs of I and Q data, while one complex signal can be constructed by one pair of I and Q data. As the value of bins of the file is 801, we can restore 1600 real signals in time domain from each frame. Here the $20th$ frame is adopted because the beam current is low and SNR is poor at this time, also it is helpful to show the advantage of modern power spectrum methods under dirty condition.

FWHM of spectrum peak is an important indicator for estimating frequency resolution, as the thinner the FWHM is, the higher frequency resolution the spectrum peak has. Also the smoothness of spectrum is another important indicator, because good smoothness is usually helpful for improving the identification of spectrum peak and expected for its grace. These three methods are compared with the spectrum peaks locating between 3.5MHz to 4MHz marked by the dotted line in Fig.3.

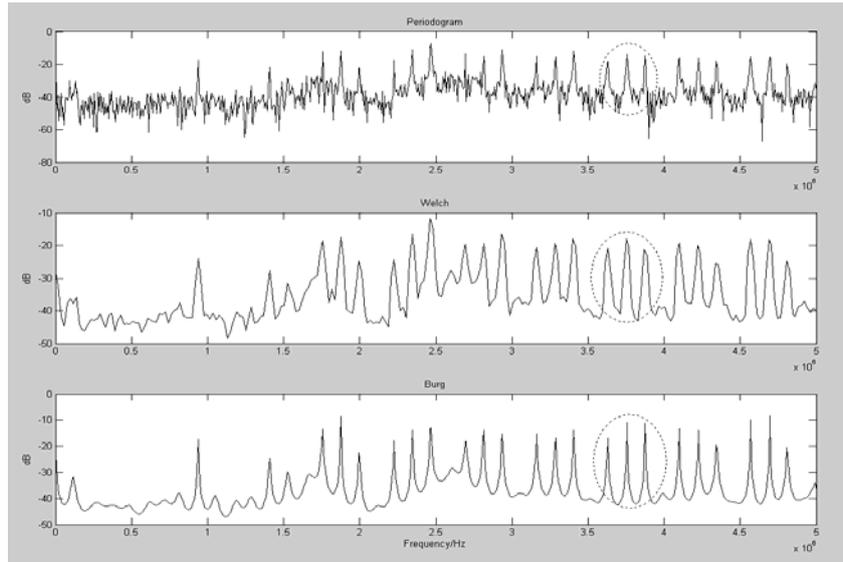

Fig.3 The performance of three power spectrum methods using IQT data

For the Welch method, hamming window is used, the length we choose is 512, and the overlap is 400. For the Burg method, 100 order parameter model is adopted. It is shown that Periodogram is less smooth than the other two methods in Fig.3, which means there are less interharmonic peaks appearing when using the other two methods. Here FWHM of lower sideband, FWHM of carrier wave and FWHM of upper sideband are calculated and compared in Table.2[12]. Welch has three larger values resulting from the decrease of data in each segment after segmenting, therefore its frequency resolution becomes lower. In other words, for the Welch method, the suppression of side lobe level is at the expense of the increase of the width of main lobe. The Burg method has three smaller values than the other two. Also the characteristic of spectrum peaks are more obvious and the summit of each peak is sharper, which is expected and helpful for tune measurement. Considering the advantages of the Burg method discussed before, the result from which is believable and available.

Table 2: The FWHM of spectrum peaks using three PSD methods

| FWHM / PSD Method | FWHM of Lower Sideband (10kHz) | FWHM of Carrier Wave (10kHz) | FWHM of Upper Sideband (10kHz) |
|---|---|---|---|
| Periodogram | 1.470 | 1.335 | 1.180 |
| Welch | 4.110 | 4.275 | 4.320 |
| Burg | 1.240 | 1.095 | 1.100 |

The complete 156 frames are shown using the Periodogram method and the Burg method In Fig.4. The result from the Burg method is on the right, from which we can see almost every group of signals of betatron tune is more obvious than the Periodogram method. In the left picture, the Periodogram method, some components are hard to see, especially for some low-frequency components, but the Burg method can still give some details of spectrum. The spectrum lines the Burg method gives look thinner and brighter, it means this method has better frequency resolution and better noise reduction, and this is useful for CSNS RCS tune measurement in the early stage as the current and SNR are low.

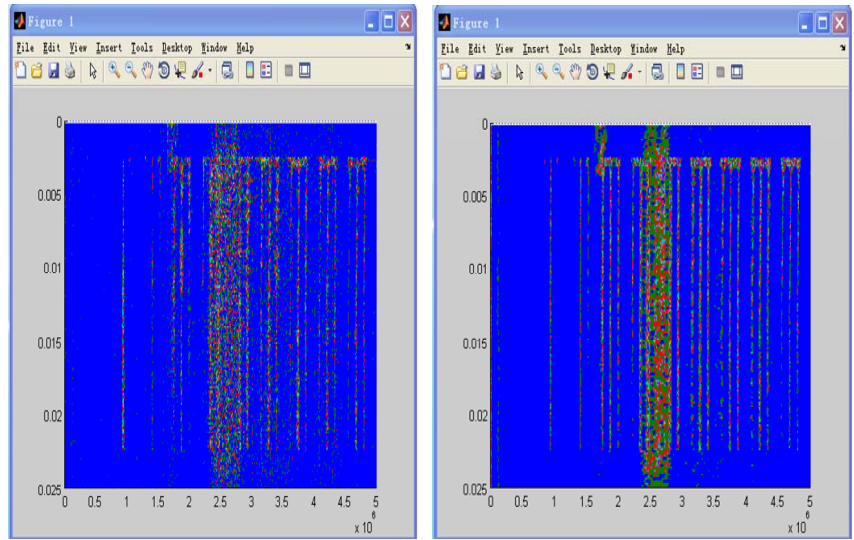
Fig.4 The power spectrum of 156 frames using Periodogram(left) and Burg(right) method

**Conclusions**

It has been shown that a modern power spectrum method such as the Burg method has higher frequency resolution than Periodogram based on Fast Fourier Transform (FFT). Usually smooth power spectrum and sharp spectrum peaks can be obtained and the accurate tune value can be calculated based on these advantages. Usually modern spectrum methods tend to produce better results than classical nonparametric methods when the available signal is relatively short or the SNR is low, it is helpful for tune measurement in CSNS RCS in the early stage.

ELECTRONICS, 2003, VOL.50, NO.3

[10] Petre Stoica, Jian Li, Hao He. Spectral Analysis of Nonuniformly Sampled Data: A New Approach Versus the Periodogram. IEEE TRANSACTIONS ON SIGNAL PROCESSING, 2009, VOL.57, NO.3

[11] http://www.tek.com/

[12] HAN Gang, LIU Xue-bin, HU Bing-liang, WANG Cai-ling. Interferogram Spectrum Reconstruction Using Modern Spectral Estimation. International Conference on Electronics, Communications and Control. Ningbo:ICECC2011. 2112-2115